\documentclass[aps,prl,twocolumn,floatfix,reprint,square,comma,numbers]{revtex4}
\usepackage{epsfig}
\usepackage{dcolumn}
\usepackage{bm}
\usepackage{color}
\usepackage{amsmath}
\usepackage[colorlinks=true,linkcolor=blue]{hyperref}

\newcommand{\bea}{\begin{eqnarray}}
\newcommand{\eea}{\end{eqnarray}}
\newcommand{\ba}{\begin{array}}
\newcommand{\ea}{\end{array}}

\begin{document}

\title{Suppression of dephasing by qubit motion in superconducting circuits}
\author{D. V. Averin$^1$}
\email{dmitri.averin@stonybrook.edu}
\author{K. Xu$^2$, Y. P. Zhong$^2$, C. Song$^2$}
\author{H. Wang$^2$}
\email{hhwang@zju.edu.cn}
\author{Siyuan Han$^3$}
\email{han@kansas.edu}
\affiliation{$^{1}$Department of Physics and Astronomy, Stony Brook University, SUNY,
Stony Brook, NY 11794-3800, USA \\
$^{2}$Department of Physics, Zhejiang University, Hangzhou, Zhejiang 310027,
China \\
$^{3}$Department of Physics and Astronomy, University of Kansas, Lawrence,
KS 66045, USA}
\date{\today }

\begin{abstract}
We suggest and demonstrate a protocol which suppresses dephasing due to the
low-frequency noise by qubit motion, i.e., transfer of the logical qubit of
information in a system of $n \geq 2$ physical qubits. The protocol requires
only the nearest-neighbor coupling and is applicable to different qubit
structures. We further analyze its effectiveness against noises with
arbitrary correlations. Our analysis, together with experiments using up to
three superconducting qubits, shows that for the realistic uncorrelated
noises, qubit motion increases the dephasing time of the logical qubit as
$\sqrt{n}$. In general, the protocol provides a diagnostic tool to measure
the noise correlations.
\end{abstract}

\pacs{  }
\maketitle

Development of superconducting qubits~\cite
{Neeley1,r2,Paik2011,Berkeley1,ucsb2011c,Fedorov2012,kelly2015} have
reached the stage when it is interesting to discuss possible architectures
of the quantum information processing circuits. The common feature of any
quantum computation process of even moderate complexity is the requirement
of information transfer between different elements of the qubit circuit.
The most straightforward way of achieving this transfer is to physically
move the quantum states representing the qubits of information along the
circuit. In the case of superconducting qubits, potential for such a direct
motion of logical qubits is offered by the so-called nSQUIDs \cite{ns1,ns2},
but operation of these circuits in the quantum regime \cite{nsq} still needs
to be demonstrated experimentally. Another method of transferring logical
qubits between different physical qubits, already developed in experiments
and adopted in this work, is based on creating controlled qubit-qubit
interaction through coupling to a common resonator
bus~\cite{sill2007,maje2007,ucsb2011c,Yang2015}. The goal of this work is to
demonstrate that, in addition to its main function, transfer of information
between different circuit elements designed to perform different functions,
have an additional notable benefit: suppression of the low-frequency dephasing.
We also show that it can be used to measure the noise correlations and, in
this way, diagnose the primary sources of the noises.

The basic mechanism of the noise suppression by qubit motion relies on the fact
that the low-frequency noise is typically produced by fluctuators - see, e.g.,
\cite{charge2004,Sendelbach2009}, in the form of impurity charges or magnetic
moments, localized in each individual physical qubit, and therefore, is not
correlated among them. Motion of a logical qubit between different physical
qubits limits the correlation time of the effective noise seen by this
qubit, and therefore suppresses its decoherence rate. This effect is
qualitatively similar to the motional narrowing of the NMR lines~\cite{kittel},
with the main difference that it is based on the controlled transfer of the
qubit state, not random thermal motion as in NMR. Also, since the effectiveness
of this mechanism is sensitive to the noise correlations not only in time, but
in space, it can be used to investigate the distribution of the primary sources
of noises in quantum circuits, promising a fast and reliable noise diagnostic
tool and, ultimately, improving the circuit performance.

Quantitatively, we start with the basic model of dephasing in a system of $n$
physical qubits, where each qubit is coupled to a source of Gaussian fluctuations
$\xi_{j}(t)$, $j=1,...,n$, of the energy difference between the computational basis
states:
\begin{eqnarray}
H_{dec} &=&-\frac{1}{2}\sum_{j=1}^{n}\sigma^{z}_{j}\xi _{j}(t)\,,  \notag \\
\langle \xi _{j}(0)\xi _{k}(t)\rangle &=&\int \frac{d\omega }{2\pi}
S_{j,k}(\omega )e^{-i\omega t}.  \label{e1}
\end{eqnarray}
Here $\sigma^{z}_{j}$ is the $z$ Pauli matrix of the $j$th qubit,
$S_{j,j}(\omega )\equiv S_{j}(\omega )$ -- spectral density of noise $\xi
_{j}(t)$ in the $j$th qubit, the terms $S_{j,k}(\omega )$, with $j\neq k$,
account for the noise correlations in different qubits, and we set $\hbar =1$.
The qubits are assumed to be free, i.e., (\ref{e1}) is the only part of
the system Hamiltonian that depends on the qubit variables.

If a logical qubit, $|\Psi \rangle =\alpha |0\rangle +\beta |1\rangle $, is
prepared at time $t=0$ as an initial state of the $j$th physical qubit and
is kept there for a period $\tau $, it will decohere due to the noise $\xi
_{j}(t)$. This decoherence process can be characterized quantitatively by
the function $F(\tau)$, defined as
\begin{equation}
F(\tau ) = \frac{\sigma _{j}(\tau )}{\sigma _{j}(0)}\,,\;\;\;\;
\sigma_{j}(\tau)= \mbox{Tr}\{\sigma ^{+}_{j} (\tau)\rho \}\,,  \label{e2}
\end{equation}
where $\rho$ is the initial density matrix of the system, which consists of
the qubit part and the part $\rho _{env}$ describing the noise source:
\begin{equation*}
\rho =|\Psi \rangle \langle \Psi |_{j}\otimes \prod_{k\neq j}|0\rangle
\langle 0|_{k}\otimes \rho _{env}\,.
\end{equation*}
Time dependence of the raising Pauli matrix, $\sigma^{+}_{j}=(\sigma^{x}_{j}
+i\sigma ^{y}_{j})/2$, of the $j$th qubit is governed by the Heisenberg
equation of motion that follows from the Hamiltonian (\ref{e1}):
$\dot{\sigma}^{+}_{j}(t)=-i\xi_{j}(t)\sigma ^{+}_{j}(t)$, and gives, as
usual,
\begin{eqnarray}
F(\tau ) &=&\langle T\exp \{-i\int_{0}^{\tau }\xi _{j}(t)dt\}\rangle \notag
\\
&=&\exp \{-\int_{0}^{\tau }dt\int_{0}^{t}dt^{\prime }\langle \xi _{j}(t)\xi
_{j}(t^{\prime })\rangle \}\,. \label{e4}
\end{eqnarray}
Here $T$ denotes the time-ordering operator, and $\langle ...\rangle$ --
averaging over the noise source $\rho_{env}$. Experimentally, the
function $F(\tau )$ is obtained by measuring the Ramsey fringes.

On the other hand, we can arrange the situation, when the logical qubit $%
|\Psi \rangle $, instead of staying just in one physical qubit for the
entire time interval $\tau$, is transferred successively from qubit $1$ to
qubit $n$ spending the time $\tau /n$ in each of them, while the transfer
processes themselves are done much faster than $\tau /n$. Such transfers can
be achieved, e.g., by applying SWAP gates to the successive pairs of
physical qubits. Then, if the transfers are done accurately, so that the
dephasing during them is negligible, the decoherence of the logical qubit $%
|\Psi \rangle$ in the total time $\tau $ is
\begin{eqnarray}
F(\tau ) &=& \exp \Big\{ -\sum_{j=1}^{n}\int_{0}^{{\tau}/{n}} dt\int_{0}^{t}
dt^{\prime }\langle \xi _{j}(t)\xi _{j}(t^{\prime})\rangle \label{e5}
\\ &-& \sum_{j<k} \int_{0}^{{\tau}/{n}}dt\int_{0}^{t} dt^{\prime
}\left\langle \xi _{k} \left(\frac{\tau}{n} (k-j)+t\right)\xi _{j}(t^{\prime
})\right\rangle \Big\}. \notag
\end{eqnarray}
If the noises are low-frequency and uncorrelated at different qubits,
decoherence is suppressed with increasing number $n$ of the physical qubits.
Indeed, in this regime, it is appropriate to neglect the quantum part of
the noise and the second sum in Eq.~(\ref{e5}) which reduces to
\begin{equation}
F(\tau )=\exp \{-\frac{1}{\pi }\int d\omega \frac{\sin ^{2}(\omega \tau /2n)}{
\omega ^{2}} \sum_{j=1}^{n}S_{j}(\omega )\}\,.  \label{e6}
\end{equation}
The low-frequency dephasing is obtained then by expanding sine in $\omega $
and keeping the first term:
\begin{equation}
F(\tau )=\exp \{-\frac{\tau ^{2}}{2 n^2}\sum_{j=1}^{n}W_{j}^{2}\}\,,\;\;%
\;W_{j}^{2}=\int_{\omega_l}^{\omega_h} \frac{d\omega }{2\pi }S_{j}(\omega
)\,.  \label{e7}
\end{equation}

For the experimentally relevant $1/f$ noise, $S_{j}(\omega)=A_{j}/|\omega |$,
the last approximation applies directly if the high-frequency cutoff of
the noise $\omega _{h}$ satisfies the condition $\tau/n \ll 1/\omega _{h}$.
As shown in the Supplementary Material~\cite{sup}, even in the opposite
regime, there are only weak logarithmic correction to scaling of the
dephasing time with $n$, and the main conclusion remains the same. The
low-frequency cutoff $\omega _{l}$ can be estimated as inverse of the time
of the experiment, and $W_{j}^{2}= (A_{j}/\pi )\ln (\omega _{h}/\omega _{l})$.
If all physical qubits have the same decoherence properties, $W_{j}=W$, we
can rewrite Eq.~(\ref{e6}) as
\begin{equation}
F(\tau)=e^{-(\tau/\tau _{d})^{2}},\;\;\;\tau _{d}=\sqrt{2n}/W\,,  \label{e8}
\end{equation}
and see that the dephasing time $\tau _{d}$ of the moving qubit increases in
comparison to the stationary qubit as $\sqrt{n}$.

If the noises at different physical qubits are correlated, one needs to take
into account both sums in Eq.~(\ref{e5}). In this case, under the same
assumptions as above, the dephasing time in Eq.~(\ref{e8}) can be written as
\begin{equation}
\frac{1}{\tau _{d}^{2}}=\frac{W^{2}}{2n^{2}}\Big[n+2\sum_{j<k}r_{j,k}\Big]\,,
\label{e9}
\end{equation}%
where the coefficient $r_{j,k}$ describe the degree of noise correlations
between the $j$th and the $k$th qubit. They are defined by the relation
$S_{j,k}(\omega )=r_{j,k}S(\omega )$, and have the property $|r_{j,k}|\leq
1$, with $r_{j,k}=1$ corresponding to full correlations, and $r_{j,k}=-1$
describing full anticorrelations \cite{dd}. Equation (\ref{e9}) shows
that if all noises are fully correlated, then $\tau _{d}=\sqrt{2}/W$, and
qubit motion does not produce any suppression of dephasing. If the noises
are completely anticorrelated between the nearest neighbors along the qubit
array, the dephasing is suppressed even more strongly than in the absence of
correlations. In this case, Eq.~(\ref{e9}) gives fully suppressed dephasing
for even $n$, and $\tau _{d}=\sqrt{2}n/W$ for odd $n$.

\begin{figure}[t]
\centering\includegraphics[width=3.4in,clip=True]{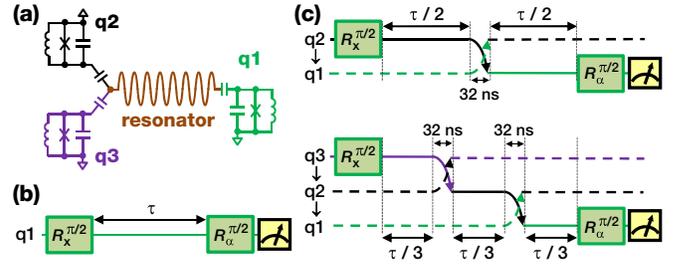}
\caption{{\protect\footnotesize {(color online) (a) Device schematic 
showing three phase qubits capacitively coupled to a central resonator. 
(b) The experimental sequence of the Ramsey fringe measurement for a single
qubit. (c) Experimental sequences for relaying the logical qubit state
between two physical qubits (top) and among three physical qubits (bottom).
The logical qubit moves along solid lines. The relay sequence starts with 
a $R_{\hat{x}}^{\protect\pi /2}$ rotation on the first qubit to create the
logical qubit state in the $x-y$ plane of the Bloch sphere, i.e., 
$|\Psi\rangle = |0\rangle -i|1\rangle $. Relaying the logical qubit state to
the next qubit is done by two successive qubit-resonator iSWAP gates that
takes 32~ns in total (symbolized by the double crossed arrows). Finally a 
$R_{\hat{\protect\alpha}}^{\protect\pi /2}$ rotation is applied to the last
physical qubit in the sequence to bring the logical qubit state to the
z-axis of the Bloch sphere for measuring the $|1\rangle$-state probability
of this qubit, $P_1$. Here $\hat{\protect\alpha}$ refers to the effective
axis that rotates in the $x-y$ plane after removing the dynamical phase, i.e.,
$\hat{\protect\alpha}=\cos (\protect\omega_{\text{R}}\protect\tau )\hat{x}%
+\sin (\protect\omega_{\text{R}}\protect\tau )\hat{y}$, where $\protect\omega%
_{\text{R}}/2\protect\pi$ is adjusted to around 25~MHz in the experiment.}}}
\label{fig1}
\end{figure}

\begin{figure}[t]
\centering\includegraphics[width=2.8in,clip=True]{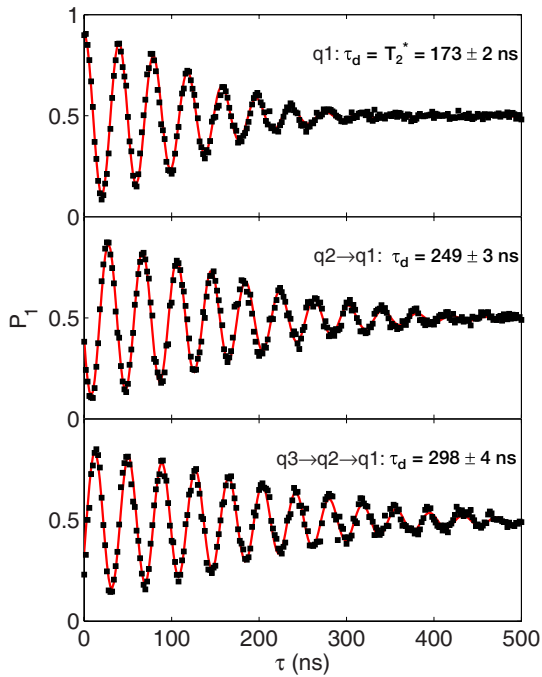}
\caption{{\protect\footnotesize {(color online) The Ramsey fringe
experimental data for sequences shown in Fig.~\protect\ref{fig1}. Red lines
are fits according to Eq.~(\protect\ref{fit}). $T_2^{\ast}$ for single qubit
can be directly compared with $\protect\tau_d$ for multiple qubits.
Statistical errors, from the measured probability spread of $\sim$2\%, are
omitted for display clarity, but are used to estimate the standard
deviations of $T_2^{\ast}$ and $\protect\tau_d$.} }}
\label{fig2}
\end{figure}

\begin{table}[tbh]
\centering
\begin{tabular}{|cc|cc|cc|}
\hline
1-qubit & $T_{2}^{*}$ (ns) & 2-qubit & ~$\tau_d$ (ns) & 3-qubit & ~$\tau_d$
(ns) \\ \hline
q1: & 173 (2) & q2$\rightarrow$q1: & ~249 (3) & q3$\rightarrow$q2$%
\rightarrow $q1: & ~298 (4) \\
q2: & 177 (1) & q3$\rightarrow$q1: & ~243 (3) & q2$\rightarrow$q3$%
\rightarrow $q1: & ~295 (4) \\
q3: & 176 (2) & q1$\rightarrow$q2: & ~245 (3) & q3$\rightarrow$q1$%
\rightarrow $q2: & ~306 (4) \\
&  & q3$\rightarrow$q2: & ~242 (3) & q1$\rightarrow$q3$\rightarrow$q2: &
~290 (5) \\
&  & q1$\rightarrow$q3: & ~244 (2) & q2$\rightarrow$q1$\rightarrow$q3: &
~298 (4) \\
&  & q2$\rightarrow$q3: & ~241 (3) & q1$\rightarrow$q2$\rightarrow$q3: &
~296 (4) \\ \hline
average: & 175.3 (2.3) & \multicolumn{2}{r|}{244.0 (3.1)} &
\multicolumn{2}{r|}{297.2 (5.5)} \\ \hline
\end{tabular}%
\caption{$T{_{2}^{\ast }}$ values for single qubit and $\protect\tau_d$
values for different experimental sequences and different qubit combinations
as outlined in Fig.~\protect\ref{fig1}. Numbers in brackets represent
standard deviations.}
\label{table1}
\end{table}

To test experimentally the mechanism of dephasing suppression by qubit motion
as discussed above, we perform the Ramsey fringe experiments (Fig.~\ref{fig1}(b))
\cite{Chiorescu2003} using up to three superconducting qubits, among which the
initial logical qubit state $|\Psi\rangle = |0\rangle -i|1\rangle $ (here
and below we ignore the normalization constant) is relayed and its phase
information is probed after the total relay time $\tau$. We use two types
of superconducting circuits in which dephasing noises differ very much in
magnitude: one features three phase qubits, each capacitively coupled to a
common resonator~\cite{Lucero2012, Zhong2014} (Fig.~\ref{fig1}(a)), and the
other one features two Xmon qubits with much reduced dephasing noises, each
as well capacitively coupled to a common resonator. The Hamiltonian of these
quantum circuits is
\begin{equation}
H=-\frac{1}{2}\sum_{j=1}^{n}\omega _{j}^{q}\,\sigma _{j}^{z}+\omega
^{r}\,a^{\dagger }a+\sum_{j=1}^{n}\lambda _{j}\,(a\sigma _{j}^{+}+a^{\dagger
}\sigma _{j}^{-}),  \label{Ham}
\end{equation}%
where the resonator frequency $\omega ^{r}$ is fixed by circuit design, the
qubit frequency $\omega _{j}^{q}$ is individually tunable, $\lambda _{j}$ ($%
\equiv \lambda$ under the homogeneous condition and $\ll \omega^r,
\omega_j^q $) describes the qubit-resonator coupling strength whose
magnitude is also fixed by circuit design, and $a^{\dagger }$ ($a$) is the
creation (annihilation) operator of the resonator field. $n$ ($=$ $1$, $2$,
or $3$) refers to the total number of physical qubits involved in each
experimental sequence.

For the phase qubit circuit, $\omega ^{r}/2\pi = 6.22$~GHz and $\lambda
/2\pi \approx 15.5$~MHz. The operation frequencies
of qubits $q1$, $q2$, and $q3$ are chosen at 5.99, 6.04, and 6.06~GHz,
respectively, for their dephasing times $T_{2}^{\ast }$s to be about the
same. Corresponding energy relaxation times $T_{1}$s are $512 \pm 6$, $538
\pm 6$, and $488 \pm 4$~ns. The dephasing times $T_{2}^{\ast }$s are $173
\pm 2$, $177 \pm 1$, and $176 \pm 2$~ns by fitting to $\ln [P_{1}(\tau
)]\propto -\tau /2T_{1}-(\tau /T_{2}^{\ast })^{2}$, where $P_1$ is the $%
|1\rangle $-state probability in the Ramsey fringe experiment~\cite{Sank2012}%
. 
Since three qubits have similar $T_{2}^{\ast }$ values, we expect that the
noise power spectral densities $S_j(\omega )$ ($j=1$, 2, and 3), which
characterize the flux-noise environments of these qubits, are approximately
at the same level~\cite{Sank2012,Yan2012,Yoshihara2014}.

At its operation frequency each qubit is effectively decoupled from the
resonator. If qubit $q1$ is in $|0\rangle -i|1\rangle $ and resonator $r$ is
in $|0\rangle $, we can turn on the qubit-resonator interaction by rapidly
matching the qubit frequency to that of the resonator for a controlled
amount of time, fulfilling an iSWAP gate~\cite{Neeley2008} to transfer the
state from $q1$ to resonator $r$, i.e., $(|0\rangle -i|1\rangle
)_{q1}|0\rangle _{r}\rightarrow |0\rangle _{q1}(|0\rangle -|1\rangle )_{r}$.
Immediately after the first iSWAP gate, we bring qubit $q2$, originally in $%
|0\rangle $, on resonance with resonator $r$ for another iSWAP gate. As
such, other than a phase factor, we effectively relay the logical qubit
state between the two qubits, i.e., $(|0\rangle -i|1\rangle )_{q1}|0\rangle
_{q2}\rightarrow |0\rangle _{q1}(|0\rangle +i|1\rangle )_{q2}$.~\cite%
{Zhong2014} For the phase qubit circuit, an iSWAP gate takes about $16$ ns
and the total time for transferring the state from one qubit to the other
qubit is about $32$ ns.

We measure the Ramsey fringe of a logical qubit which spends equal amount of
time in each of the $n \geq 2$ physical qubits. The sequences are illustrated 
in Fig.~\ref{fig1}(c). The obtained Ramsey fringe is fitted according 
to~\cite{Sank2012},
\begin{equation}
P_{1}(\tau )=A\exp \left[ -\frac{\tau }{2T_1^{ave}}-\left( \frac{\tau }{\tau
_{d}}\right) ^{2}\right] \cos (\omega_{\text{R}}\tau +B)+ C,  \label{fit}
\end{equation}%
where $T_{1}^{ave}$ is fixed as the average of all qubits involved and $%
\tau_d$ is the effective dephasing time for the logical qubit as obtained
from the fit (so are the constants A, B, C, and $\omega_{\text{R}}$). 
Representative experimental data and fitting curves are shown in 
Fig.~\ref{fig2}.

Controlled motion of the logic qubit are attempted under various
experimental conditions. Table~\ref{table1} lists $\tau _{d}$ values of the
logic qubit obtained using different experimental sequences and different
qubit combinations. The Ramsey fringe measurements using two qubits (three
qubits) show that dephasing times of the logic qubits are extended to about $%
244.0 \pm 3.1$ ns ($297.2 \pm 5.5$ ns), averaged a gain by a factor of $%
1.392=0.984\sqrt{2}$ ($1.695=0.979\sqrt{3}$) compared with those from the
single-qubit measurements ($175.3 \pm 2.3$ ns). As expected from Eq.~(\ref%
{e6}), the logic qubit dephasing time $\tau _{d}$ scales very well with the
square root of the number of physical qubits, $\sqrt{n}$. The similar
scaling is also observed using two Xmon qubits, where the single-qubit $%
T_2^{\ast}$ values are about $1$~$\mu$s, achieving a gain of $1.405=0.993%
\sqrt{2}$ (Ramsey fringe data not shown). Our result clearly demonstrates
that dephasing caused by uncorrelated low-frequency noises can be reduced by
a factor of $\sqrt{n}$ by moving the logic qubit state along an array of $%
n\geq 2$ physical qubits.

The result demonstrated above is based on the fact that noises at different 
qubits were completely uncorrelated. In general, degree of the noise 
suppression by qubit motion method depends on the noise correlation magnitude.
Assuming that the noise environments $S_j(\omega )$, $j = 1$ and 2, for two
qubits are at the same level, Eq.~(\ref{e9}) is reduced to
\begin{equation}
\frac{1}{\tau _{d}^{2}}=\frac{W^{2}}{4}\left( 1+r_c\right) ,\;\ \ \tau _{d}=
\sqrt{\frac{2}{1+r_c}}T_{2}^{\ast },  \label{e12}
\end{equation}%
where $T_{2}^{\ast }=\sqrt{2}/W$ and $r_c \equiv r_{1,2}$. The monotonous
dependence of $\tau _{d}$ on the correlation coefficient $r_c$ of the
two-qubit noises provides a much needed guide for measuring noise
correlations. Since the Ramsey fringe measurement is much faster than the
conventional two-point correlation measurement~\cite{Bialczak2007,Yan2012},
the latter of which usually takes at least a few hours in order to cover a
wide range in spectrum, the mechanism of qubit motion may provide a fast and
reliable diagnostic tool for identifying the primary sources of noises in
complex quantum circuits.

\begin{figure}[t]
\includegraphics[width=2.7in]{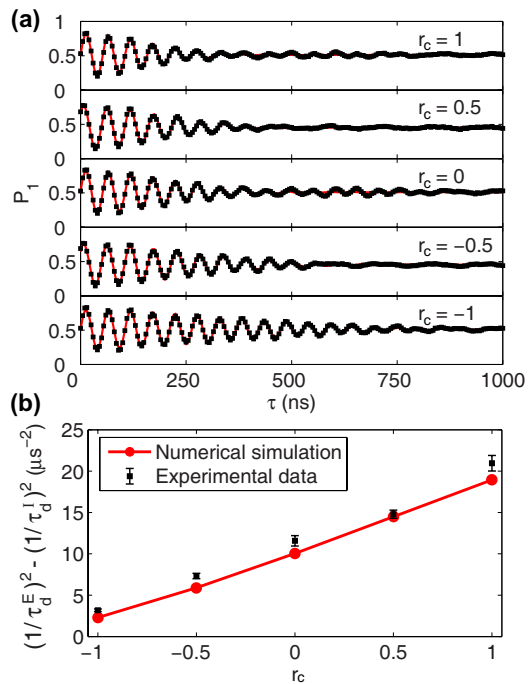}\newline
\caption{(a) The 2-qubit Ramsey fringe results under artificial noises with
different correlations $r_c$ as indicated. Dots are experimental data and
lines are fits. Fitted $\protect\tau_d^E$ values are $213 \pm 4$, $251 \pm 4$,
$281 \pm 7$, $345 \pm 6$, and $485 \pm 8$~ns from top to bottom.
(b) $(1/\protect\tau_d^E)^{2}-(1/\protect\tau_d^I)^{2}$ versus $r_c$ (black
squares), where $\protect\tau_d^I$ is the dephasing time from the 2-qubit
Ramsey fringe experiment under no artificial noise. Error bars are estimated based
on uncertainties of $\tau_d^E$ and $\tau_d^I$. The numerical result (red dots) is
obtained by solving the Schrodinger-Langevin equation \cite{Yasue1978} with a
time-dependent Hamiltonian. Line is a guide to the eye.}
\label{fig3}\centering
\end{figure}

We experimentally emulate the monotonous dependence of $\tau _{d}$ on $r_c$
in Eq.~(\ref{e12}) using two Xmon qubits, where the much reduced intrinsic
dephasing noises make it easier to inject controllable noises~\cite%
{Biercuk2009}. Here the intrinsic noises refer to any noises associated with
the device or measurement setup, in contrast to the extrinsic ones that
specifically refer to our controlled noises.
We first set the operation frequencies of the two Xmon qubits, exposing them
to the same level of intrinsic noise environments, characterized by $%
S_j^{I}(\omega)$, $j=1$ and 2. We then apply strong low-frequency noises,
digitally synthesized with an adjustable correlation coefficient $r_c$, to
the two qubits so that both qubits' dephasing rates are dominated by these
extrinsic noises. It is verified that for each qubit $T_{2}^{\ast }$ is
reduced to about $220$~ns due to the combination of the noise powers of $%
S_j^I(\omega)$ and $S_j^E(\omega)$, $j=1$ or 2, where $S_j^E(\omega)$
characterizes the synthesized noise power spectral densities (see
Supplemental Material~\cite{sup}). Synthesized noises are simultaneously
applied during the 2-qubit Ramsey fringe experiments. Resulted Ramsey
fringes shown in Fig.~\ref{fig3}(a) can be used to estimate $\tau_d^E$, the
logic-qubit dephasing time determined by both $S_j^E(\omega)$ and $S_j^I(\omega)$, $j = 1$ and 2. It
is observed that $\tau_{d}^E$ increases monotonically when the correlation
coefficient $r_c$ changes from $1$ (perfectly correlated) to $-1$
(anti-correlated), in agreement with Eq.~(\ref{e12}). In Fig.~\ref{fig3}(b)
we plot $(1/\tau_d^E)^{2}-(1/\tau_d^I)^{2}$ versus $r_c$ (black squares with 
error bars), where $\tau_d^I$ is the logic-qubit dephasing time dominated 
only by $S_j^I(\omega)$, $j = 1$ and 2, as measured with the 2-qubit Ramsey 
sequence under no extrinsic noises. Also shown in Fig.~\ref{fig3}(b) are the
numerical simulation results. The experimental and simulation data are slightly 
different from the prediction by Eq.~(\ref{e12}), likely due to the fact that 
experimentally synthesized extrinsic noises only cover down to 10 kHz in 
spectrum as limited by hardware resource (see Supplemental Material~\cite{sup}), 
while Eq.~(12) works better for lower-frequency noises.

To summarize, we propose a new scheme to suppress low-frequency induced
dephasing of logic qubit states by moving the quantum information along an
array of $n \geq 2$ physical qubits. We have shown that in general qubit
motion can make dephasing time $\tau _{d}$ of the logic qubits longer than
that of the physical qubits $T_{2}^{\ast },$ as long as noises on physical
qubits are not completely correlated. For uncorrelated noises, our model
predicts a simple scaling $\tau _{d}=\sqrt{n}T_{2}^{\ast }$. We have
experimentally implemented the qubit motion scheme to suppress dephasing
using superconducting circuits with up to $n=3$ Josephson qubits. The
results agree very well with that expected from the model. Furthermore,
using synthesized noises we have demonstrated that measuring the ratio $\tau
_{d}/T_{2}^{\ast }$ allows one to determine quantitatively the correlation
coefficient $r_c$ between noises on two physical qubits, which is difficult
to obtain with conventional methods. The qubit motion method can be readily
applied to more qubits to further suppress dephasing and it is
straightforward to incorporate qubit motion with quantum gate operations on
logic qubits. Our results thus open a new venue for improving performance of
logic qubits and gaining insight on low-frequency noises in complex quantum
information processing circuits.\newline

The authors thank John Martinis for providing the devices used in the
experiment. This work was supported by the National Basic Research Program
of China (2014CB921200, 2012CB927404), US NSF grants PHY-1314758 and
PHY-1314861, the National Natural Science Foundation of China (11434008,
11222437), and Zhejiang Provincial Natural Science Foundation of China
(LR12A04001).\newline

\end{document}